\documentclass[aps,prb,superscriptaddress,twocolumn]{revtex4}
\usepackage{amssymb}
\usepackage{graphicx}
\usepackage{amsmath}
\usepackage[colorlinks=true,linkcolor=blue,citecolor=blue,urlcolor=blue]{hyperref}

\newcommand{\be}{\begin{equation}}
\newcommand{\ee}{\end{equation}}

\newcommand{\bs}{\begin{split}}
\newcommand{\es}{\end{split}}

\usepackage[normalem]{ulem}
\usepackage{amsmath,amssymb}
\usepackage{multirow}
\usepackage{xcolor,soul}
\usepackage{srcltx}
\usepackage{hyperref,graphicx}

\begin{document}

\title{Resonances, mobility edges and gap-protected Anderson localization in generalized disordered mosaic lattices}
\author{Stefano Longhi}
\thanks{stefano.longhi@polimi.it}
\affiliation{Dipartimento di Fisica, Politecnico di Milano, Piazza L. da Vinci 32, I-20133 Milano, Italy}
\affiliation{IFISC (UIB-CSIC), Instituto de Fisica Interdisciplinar y Sistemas Complejos, E-07122 Palma de Mallorca, Spain}

\begin{abstract}
Mosaic lattice models have been recently introduced as a special class of disordered systems displaying resonance energies, multiple mobility edges and anomalous transport properties.  In such systems on-site potential disorder, either uncorrelated or incommensurate,  is introduced solely at every equally-spaced sites within the lattice, with a spacing $M \geq 2$. A remarkable property of disordered mosaic lattices is the persistence of extended states at some resonance frequencies that prevent complete Anderson localization, even in the strong disorder regime. Here we introduce a broader class of mosaic lattices and derive general expressions of mobility edges and localization length for incommensurate sinusoidal disorder, which generalize previous results [Y. Wang {\it et al.}, Phys. Rev. Lett. {\bf 125}, 196604 (2020)]. 
For both incommensurate and uncorrelated disorder, we prove that Anderson localization is protected by the open gaps of the disorder-free lattice, and derive some general criteria for complete Anderson localization. The results are illustrated by considering a few models, such as the mosaic Su-Schrieffer-Heeger (SSH) model and the trimer mosaic lattice. 
 \end{abstract}

\maketitle

\section{Introduction}

Anderson localization and mobility edges are fundamental concepts in the study of disordered systems  \cite{R1,R2,R2bis,R2b,R2c,R3}, playing a pivotal role in different areas of physics, ranging  from condensed matter physics \cite{R2,R2b,R2c,R3} to ultracold atoms \cite{R4,R5,R6,R7,R8,R8b,R9,R10,R11,R11b} and disordered photonics \cite{R12,R13,R14,R15,R16,R16b,R16c}. 
In his seminal article \cite{R1}, Anderson predicted that in one-dimensional or two-dimensional
disordered systems with uncorrelated disorder all states are localized at any disorder strength, while a disorder threshold is required for localization in three-dimensional systems \cite{R16d}.
However, in certain one-dimensional systems with correlated disorder or with an incommensurate potential \cite{R16e,R16f,S1,S1b,S1c,S2,S3}, a disorder threshold can arise \cite{R16e,R16f}, and extended (ergodic or non-ergodic) states can coexist with localized ones \cite{S2}, separated
by mobility edges \cite{R2} in the energy spectrum. The study of mobility edges, metal-insulator phase transitions, and critical states in such systems has generated significant interest, resulting in a substantial body of literature on the subject (see e.g. \cite{S2,S3,S3b,S4,S4b,S5,S6,S6b,S7,S8,S9,S10,S11,S12,S13,S14,S15,S16,S17,S18,S19,S20,S21,S22,S23,S24,S25,S26,S27,S28,S29,S30,S31} and references therein).\\ 
Recently, great attention has been devoted to studying the localization properties of mosaic lattices with incommensurate or random (uncorrelated) disorder \cite{M1,M1b,M2b,M2c,M2,M3,M4,M5,M6,M7,M8,M9,M10,M11,M12,M13} --also referred to as diluted or periodic Anderson model in earlier works \cite{D1,D2,D3,D4,D5} -- where onsite potentials are introduced only at equally spaced sites within the lattice. For an incommensurate potential, mosaic lattices provide new type of quasiperiodic
systems exhibiting exact mobility edges \cite{M1} (see also \cite{M4,M5,M13}), while for uncorrelated disorder (diluted Anderson models \cite{M2,M3,D1,D2,D3,D4,D5}) or for other types of mosaic potentials, such as Wannier-Stark or unbounded potentials \cite{
M8,M9,M10,M11,M12}, 
extended or weakly localized states with diverging localization length emerge near some resonance  energies (or quasi-resonances in finite-sized lattices) \cite{M3,D1,D2,D3}, leading to 
delocalization and anomalous dynamical behaviors \cite{M2,M3,M10,D4,M13}. The appearance of resonances is analogous to the one found in other models with correlated disorder, such as in the famous random dimer model \cite{S1,S1b}, and persists at any disorder strength, preventing Anderson localization of the entire eigenstates. Basically, the wave functions of the extended states at the resonance energies in the infinitely-extended mosaic lattice are dark at the equally-spaced sites where the disordered potential is applied, and thus they are insensitive to both type and strength of disorder. Previous studies have been focused to
some specific mosaic lattice models, lacking for some generality of the results. For example, for incommensurate mosaic potentials the derivation of exact mobility edges and localization length (inverse of Lyapunov exponent) have been limited to lattices with nearest-neighbor hopping and used Avila's global theory, which could be of nontrivial application in generic mosaic lattice models or even not applicable \cite{M9}. Finally, a rather open question, which has not been addressed in previous studies, is whether resonance-induced delocalization in mosaic lattices with incommensurate or uncorrelated disorder can be prevented and complete Anderson localization restored, at least in the strong disorder regime.\par
In this work we consider a broad class of one-dimensional disordered mosaic lattices, with either incommensurate or uncorrelated disorder, beyond models so far investigated \cite{M1,M2,M3,M4,M5,
M7,M8,M9,M10,M11,D1,D2,D3,D4}. For incommensurate sinusoidal disorder (Aubry-Andr\'e or almost Mathieu mosaic lattices), we provide general analytical form of mobility edges and inverse localization length, without resorting to Avila's global theory. For uncorrelated  disorder, a general asymptotic form of Lyapunov exponent is derived near resonance energies, and exact results are given for the class of mosaic Lloyd models. Finally, general criteria are given to avoid the appearance of resonance energies in the energy spectrum, thus restoring complete Anderson localization of all eigenstates. For incommensurate disorder, it is shown that a sufficient condition for complete Anderson localization at strong enough disorder is that all the gaps of the disorder-free superlattice are open and  the resonance energies fall deeply insight into the gaps. For random disorder, complete Anderson localization is obtained at any disorder strength if and only all the gaps are open. Shortly, open gaps of the disorder-free superlattice provide a protection of Anderson localization, while gap closing weakens Anderson localization owing to the appearance of extended states at resonance energies in the spectrum. The results are illustrated by considering a few models, such as the mosaic 
Su-Schrieffer-Heeger (SSH) model  and the trimer mosaic lattice.

\begin{figure}[t]
   \centering
    \includegraphics[width=0.5\textwidth]{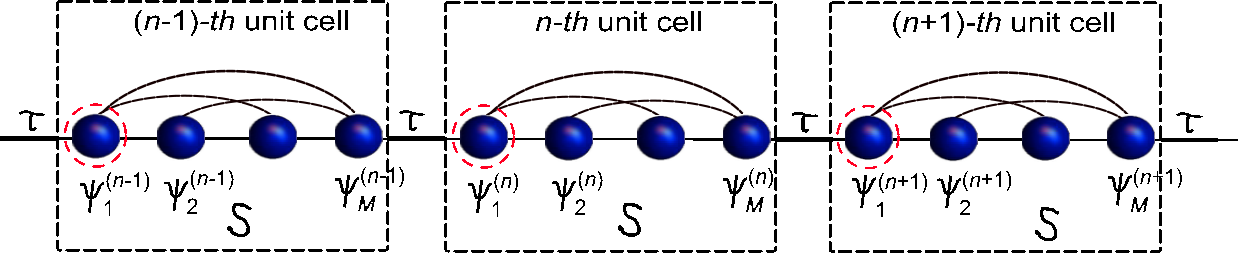}
   \caption{Schematic of a disordered mosaic lattice with inter-cell hopping amplitude $\tau$ and intra-cell Hamiltonian $\mathcal{S}$. On-site potential disorder $V_n$ is applied every $M$ sites in the lattice ($M=4$ in the figure), indicated by the red dashed circles. In the disorder-free limit ($V_n=0$) one has a superlattice, where  each unit cell comprises $M$ sites with rather arbitrary on-site energies and intra-cell hopping amplitudes, described by the intra-cell Hamiltonian $\mathcal{S}$.}
    \label{fig1}
\end{figure}

\section {Model,  energy spectrum and resonance energies} 
\subsection{Model}
 We consider a one-dimensional tight-binding lattice with mosaic disorder  described by a tight-binding Hamiltonian $H$, where on-site potential is introduced solely at every $M$ sites of the lattice \cite{M1}. We assume that the disorder-free (clean) lattice, described by the Hamiltonian $H_0$, displays discrete translation invariance for a spatial displacement by $M$, so that it can be regarded as a superlattice comprising $M$ sublattices and each unit cell contains $M$ sites, as schematically shown in Fig.1. Let us indicate by $\psi_n^{ ( \alpha)}$ the wave amplitude at the $\alpha$-th site of the $n$-th unit cell of the superlattice, where $\alpha=1,2,...,M$  is the sublattice index and $n=0, \pm 1, \pm 2 , \pm 3,...$ is the unit cell index. A mosaic lattice is obtained from the disorder-free model by introducing an on-site potential $V_n$ at the sites of the sublattice $\alpha=1$ solely. 
  The spectral problem for the most general mosaic lattice of Hamiltonian $H$ can be written as
 \begin{equation}
 E \psi_{n}^{(\alpha)}= \sum_{l=-\infty}^
{\infty} \sum_{\beta=1}^M S_{\alpha, \beta}^
{(n-l)} \psi_{l}^{(\beta)}+V_n \delta_{\alpha,1} \psi_{n}^{(1)}  \label{eq1}
 \end{equation}
 where  $S_{\alpha, \beta}^{(n-l)}=S_{\beta, \alpha}^{(l-n) *}$ describe the on-site energies (for $n=l$ and $\alpha=\beta$), inter-hopping (for $n \neq l$) and intra-hopping (for $n=l$) amplitudes among different sites of the clean superlattice, and $V_n$ is the applied mosaic disorder. We focus our attention to the family of superlattices where inter-hopping occurs solely between adjacent sites of the nearest neighbor unit cells, as schematically shown in Fig.1, while no restrictions are made for the Hermitian matrix $\mathcal{S} \equiv S_{\alpha,\beta}^{(0)}=S_{\beta,\alpha}^{(0)*}$ describing intra-hopping processes and on-site energies of the clean lattice (intra-cell Hamiltonian). Indicating by $\tau$ the inter-hopping amplitude, for $ l \neq n$ one has $S^{(n-l)}_{\alpha,\beta}=\tau ( \delta_{l,n-1} \delta_{\alpha,1} \delta_{\beta_M}+\delta_{l,n+1} \delta_{\alpha,M} \delta_{\beta,1})$ and thus the spectral problem defined by Eq.(\ref{eq1}) reads explicitly
 \begin{equation}
 E 
 \left(
 \begin{array}{c}
 \psi_{n}^{(1)} \\
 \psi_{n}^{(2)} \\ 
 ... \\
 \psi_{n}^{(M-1)} \\
 \psi_{n}^{(M)}
 \end{array}
 \right)=
 \mathcal{S}
 \left(
 \begin{array}{c}
 \psi_{n}^{(1)} \\
 \psi_{n}^{(2)} \\ 
 ... \\
 \psi_{n}^{(M-1)} \\
 \psi_{n}^{(M)}
 \end{array}
 \right)+
 \left(
 \begin{array}{c}
 V_n \psi_{n}^{(1)} +\tau \psi_{n-1}^{(M)}\\
0  \\ 
 ... \\
 0 \\
\tau \psi_{n+1}^{(1)}
 \end{array}
 \right) \label{eq2}
 \end{equation}
 where $n=0, \pm 1, \pm2, \pm 3,...$ is the unit cell index.\\
 The disorder $V_n$ is assumed to be either an incommensurate potential, such as the sinusoidal potential 
 \begin{equation}
 V_n=2V \cos (2 \pi \alpha n + \theta) \label{poten}
 \end{equation}
 with amplitude $V$ and irrational Diophantine frequency $ \alpha$, describing the class of Aubry-Andr\'e (or almost Mathieu) mosaic lattices \cite{
M1,M4,M5,M7,M11,M13}, 
 or independent random variables with the same probability density distribution $f(V)$, describing the class of Anderson mosaic lattices (diluted Anderson model \cite{M2,M3,D1,D2,D3,D4,D5}).
 
 \subsection{Spectrum of the disorder-free lattice}
 We indicate by $\sigma(H_0)$ the energy spectrum of the disorder-free lattice, obtained by assuming $V_n=0$.
Clearly, owing to discrete translational invariance $\sigma(H_0)$ is absolutely continuous and composed by $M$ energy bands, with corresponding extended eigenfunctions of Bloch type. After letting
 \begin{equation}
 \left(
 \begin{array}{c}
 \psi_{n}^{(1)} \\
 \psi_{n}^{(2)} \\ 
 ... \\
 \psi_{n}^{(M-1)} \\
 \psi_{n}^{(M)}
 \end{array}
 \right)=
 \left(
 \begin{array}{c}
 A^{(1)} \\
 A^{(2)} \\ 
 ... \\
 A_{n}^{(M-1)} \\
 A_{n}^{(M)} 
 \end{array}
 \right) \exp(i qn),
 \end{equation}
 where $-\pi < q \leq \pi$ is the Bloch wave number varying in the first Brillouin zone, the dispersion curves $\mathcal{E}_{\lambda}(q)$ of the $M$ superlattice minibands are the eigenvalues of the following $M \times M$ matrix
 \begin{widetext}
 \begin{equation}
 \mathcal{A}=\left(
 \begin{array}{cccccc}
S_{1,1} & S_{1,2} & S_{1,3} & ... & S_{1,M-1} & S_{1,M} +\tau \exp(-iq) \\
S_{2,1} & S_{2,2} & S_{2,3} & ... & S_{2,M-1} & S_{2,M} \\
... & ... & ...  & ... & ... & ... \\
S_{M-1,1} & S_{M-1,2} & S_{M-1,3} & ... & S_{M-1,M-1} & S_{M-1,M} \\
S_{M,1} + \tau \exp(iq) & S_{M,2} & S_{M,3} & ... & S_{M,M-1} & S_{M,M}   
 \end{array}
 \right) \label{matrixA}
 \end{equation}
 \end{widetext}
 as $q$ varies in the first Brillouin zone, ordered such that
 \begin{equation}
 \mathcal{E}_{1}(q) \leq  \mathcal{E}_{2}(q) \leq ...  \leq \mathcal{E}_{M-1}(q) \leq  \mathcal{E}_{M}(q). 
 \end{equation}
 The gap between adjacent bands $\lambda$ and $(\lambda+1)$ closes when for some $q=q_0$ one has $\mathcal{E}_{\lambda}(q_0)=\mathcal{E}_{\lambda+1}(q_0)$. In the absence of gauge fields, i.e. whenever $\mathcal{S}$ is a real symmetric matrix, gap closing occurs at the band center or band edges, i.e. at $q_0=0$ or $q_0= \pm \pi$. In the following analysis, we will typically assume real hopping amplitudes, so that $\mathcal{S}$ is a real and symmetric matrix. 
 
 \subsection{Spectrum of the disordered mosaic lattice}
 Let us indicate by $\sigma(H)$ the energy spectrum of the disordered mosaic lattice. To determine $\sigma(H)$, we extend the block decimation scheme earlier introduced in \cite{D1,D2}. To this aim, let us assume that $E$ is not an eigenvalue of $\mathcal{S}$ \cite{footzero} and consider the Hermitian matrix
 \begin{equation}
 \mathcal{K}(E)=(E-\mathcal{S})^{-1}. \label{eq3}
 \end{equation}
Note that, when $\mathcal{S}$ is real symmetric, $\mathcal{K}(E)$ is real symmetric as well for real energies $E$.
 Equations (\ref{eq2}) can be formally solved, yielding for $\psi_n^{(1)}$ and $\psi_n^{(M)}$ the following set of coupled discrete equations
 \begin{eqnarray}
 \psi_n^{(1)} & = & \mathcal{K}_{11}(E) \left( V_n \psi_{n}^{(1)} +\tau \psi_{n-1}^{(M)}\right)+ \tau \mathcal{K}_{1M}(E) \psi_{n+1}^{(1)} \;\;\; \label{eq4} \\
 \psi_n^{(M)} & = & \mathcal{K}_{M1}(E) \left( V_n \psi_{n}^{(1)} +\tau \psi_{n-1}^{(M)}\right)+ \tau \mathcal{K}_{MM}(E) \psi_{n+1}^{(1)}. \;\;\;\;\;\; \label{eq5}
 \end{eqnarray}
 For $\mathcal{K}_{11}(E) \neq 0$, one can eliminate the amplitudes $\psi_n^{(M)}$ from Eqs.(\ref{eq4}) and (\ref{eq5}), yielding the following second-order difference equation for the ampitudes $\psi_n^{(1)} $
 \begin{equation}
 \tau \mathcal{K}_{1M} \psi_{n+1}^{(1)}+\tau \mathcal{K}_{M1} \psi_{n-1}^{(1)}+\mathcal{K}_{11} V_n  \psi_{n}^{(1)}=\mathcal{E}  \psi_{n}^{(1)} \label{eq6}
 \end{equation}
 where we have set
 \begin{equation}
 \mathcal{E}=1+\tau^2 |\mathcal{K}_{1M}|^2-\tau^2 \mathcal{K}_{11} \mathcal{K}_{MM}. \label{eq7}
 \end{equation}
 Formally, Eq.(\ref{eq6}) can be regarded as the spectral problem of a disordered one-dimensional tight-binding lattice with nearest-neighbor { \em energy-dependent} hopping amplitude $\tau \mathcal{K}_{1M}(E)=\tau \mathcal{K}_{M1}^*(E)$  and {\em energy-dependent} on-site potential $W_n=\mathcal{K}_{11}(E) V_n$. This reduction is extremely useful to provide general and sometimes analytical results on the energy spectrum $\sigma(H)$ and localization properties (i.e. Lypaunov exponent $\gamma(E)$) of the mosaic lattice once they are known for the simpler nearest-neighbor tight-binding lattice with on-site disorder described by Eqs.(\ref{eq6}) and (\ref{eq7}).\\
 The localization properties of the mosaic lattice are greatly affected by the appearance of {\em resonance energies}, which are defined as the roots of the equation
 \begin{equation} 
 \mathcal{K}_{11}(E)=0 \label{roots}
 \end{equation}
 in correspondence of which the effective disorder term $W_n=\mathcal{K}_{11}(E) V_n$ in Eq.(\ref{eq6}) vanishes. 
 The explicit expression of the matrix element $\mathcal{K}_{11}(E)=\{ (E-\mathcal{S})^{-1}\}_{11}$ reads
 \begin{equation}
 \mathcal{K}_{11}(E)=\frac{{\rm det} (E-\mathcal{B})}{{\rm det} (E-\mathcal{S})}
 \end{equation}
where $\mathcal{B}$ is the principal submatrix of $\mathcal{S}$, obtained from $\mathcal{S}$ by deleting the first row and the first column, i.e.
 \begin{equation}
 \mathcal{B}=\left(
 \begin{array}{cccc}
 S_{2,2} & S_{2,3} & ... & S_{2,M} \\
 S_{3,2} & S_{3,3} & ... & S_{3,M} \\
... & ....  & ... & ... \\
 S_{M,2} & S_{M,3} & ... & S_{M,M} \\
  \end{array}
 \right) \label{matrixB}
 \end{equation}
  Hence, the roots of the equation (\ref{roots}), i.e. the resonance energies, are the $(M-1)$ real eigenvalues  $E_{1}$, $E_{2}$,..., $E_{M-1}$ of the Hermitian principal submatrix $\mathcal{B}$ of $\mathcal{S}$. It can be readily shown that they are interlaced with the minibands of the disorder-free superlattice, namely one has
 \begin{equation}
 \mathcal{E}_1(q) \leq E_1 \leq  \mathcal{E}_2(q) \leq E_3 \leq ... \leq \mathcal{E}_M(q) \leq E_{M-1} \leq  \mathcal{E}_M(q). \label{ineq}
 \end{equation}
 Such as result basically follows from the Cauchy's interlace theorem for eigenvalues of Hermitian matrices (see e.g. \cite{inter}) and the fact that the matrix $\mathcal{B}$ is also the principal submatrix of $\mathcal{A}$ (in fact deleting the first raw and first column from either matrices $\mathcal{S}$ or $\mathcal{A}$ yields the same principal submatrix $\mathcal{B}$).

 When the energy $E$ in the spectral problem Eq.(\ref{eq1}) is set equal to one of such resonance energies $E_{\lambda}$, according to Eqs.(\ref{eq4}) and (\ref{eq5}) the spectral problem is readily solved by letting
 \begin{equation}
 \psi_n^{(1)}=0 \; , \; \;   \psi_n^{(M)} =  \mathcal{K}_{M1}(E_{\lambda}) \tau \psi_{n-1}^{(M)} \label{eq8}
 \end{equation} 
 corresponding to a wave function which is dark (i.e. vanishing) in the entire sublattice $\alpha=1$, where the on-site potential disorder is applied. Clearly, from Eq.(\ref{eq8}) the boundedness condition for 
 $\psi_n^{(M)}$ as $ n \rightarrow \pm \infty$ 
indicates that the resonance energy $E={E}_{\lambda}$ ($\lambda=1,2,...,M-1$)  belongs to the energy spectrum $\sigma(H)$ if and only if $|\mathcal{K}_{M1}(E_{\lambda}) \tau |=1$, corresponding to a vanishing Lyapunov exponent $\gamma(E_{\lambda})=0$. In other words, a resonance energy $E_{\lambda} \in \sigma(H)$ if and only if $|\mathcal{K}_{M1}(E_{\lambda}) \tau |=1$, and in this case the corresponding eigenstate is extended and independent of the type and strength of disorder, so that $E_{\lambda} \in \sigma(H_0)$ as well.\\ 
Clearly, the energy spectrum $\sigma(H)$ and localization properties of the disordered mosaic lattice depend on the specific model and the type of disorder, and can be obtained using the reduced model [Eqs.(\ref{eq6}) and (\ref{eq7})]. Let us consider the two important cases of incommensurate (sinusoidal) mosaic potential and uncorrelated (Anderson) mosaic disorder.
\subsubsection{Incommensurate (sinusoidal) disorder} 
 Let us assume that $V_n$ describes the incommensurate (almost Mathieu) potential, given by Eq.(\ref{poten}), 
 with amplitude $V \geq 0$ and irrational Diophantine frequency $ \alpha$. According to Eq.(\ref{eq6}), the spectral problem of the mosaic lattice is reduced to the one of the ordinary Aubry-Andr\'e model \cite{R16e,R16f}, and thus the following results hold:\\
 (i) For any energy $E \in \sigma(H)$ such that
 \begin{equation}
 V> \tau \left|  \frac{\mathcal{K}_{1M}(E)}{\mathcal{K}_{11}(E)} \right| \label{uff1}
\end{equation}
 the eigenstate is exponentially localized with an inverse localization length (Lypaunov exponent) given by
 \begin{equation}
 \gamma(E)= {\rm log} \left|  \frac{\mathcal{K}_{11}(E) V }{\tau \mathcal{K}_{1M}(E)} \right|. \label{uff2}
\end{equation}
(ii) For any energy $E \in \sigma(H)$  such that 
  \begin{equation}
 V< \tau \left|  \frac{\mathcal{K}_{1M}(E)}{\mathcal{K}_{11}(E)} \right| \label{extended}
\end{equation}
 the eigenstate is extended and ergodic.\\ 
 (iii) Mobility edges, separating exponentially-localized eigenstates and extended states, arise at the energies satisfying the condition
   \begin{equation}
 V= \tau \left|  \frac{\mathcal{K}_{1M}(E)}{\mathcal{K}_{11}(E)} \right|. \label{mobility}
\end{equation}
At such energies, when $E \in \sigma(H)$ the corresponding eigenstates are critical.\\

As illustrative examples, let us specify the general analysis to three models.\\
1) {\it Mosaic lattice with homogeneous nearest-neighbor hopping}. This is the simplest mosaic lattice model, which was earlier introduced in Ref.\cite{M1} as an example of a one-dimensional lattice dispalying exact mobility edges. Here we show that the same results of Ref.\cite{M1} are obtained as a special case of our general analysis, without resorting to Avila's global theory. The $M \times M$ matrix $\mathcal{S}$ for this model is given by the tridiagonal Jacobi matrix
\begin{equation}
\mathcal{S}=\left(
\begin{array}{ccccccc}
0 & \tau & 0 & ... & 0 & 0 & 0 \\
\tau & 0 & \tau & ... & 0 & 0 & 0\\
... & ... & ... & ... & ... & ... & ... \\
0 & 0 & 0 & ... & \tau & 0 & \tau \\
0 & 0 & 0 & ... & 0 & \tau & 0
\end{array}
\right)
\end{equation}
where $\tau$ is the nearest-neighbor hopping amplitude. The matrix $\mathcal{K}(E)=(E-\mathcal{S})^{-1}$ is then readily obtained from the inverse of a Joacobi tridiagonal matrix \cite{Jacobi},  and the elements $\mathcal{K}_{11}(E)$ and $\mathcal{K}_{1M}(E)$ read 
\begin{equation}
\mathcal{K}_{11}(E)= \frac{\sinh (M \lambda)}{\sinh [(M+1) \lambda]} \; , \;  \mathcal{K}_{1M}(E)= \frac{\sinh ( \lambda)}{\sinh [(M+1) \lambda]}
\end{equation}
where we have set
\begin{equation}
\cosh \lambda \equiv \frac{E}{2 \tau}
\end{equation}
From the above relations one has
\begin{equation}
\frac{\mathcal{K}_{1M}}{\mathcal{K}_{11}}= \frac{\exp(\lambda)-\exp(-\lambda)}{\exp(M \lambda)-\exp(-M \lambda)}. \label{uff3}
\end{equation}
Therefore, from Eqs.(\ref{mobility}) and \ref{uff3}) it follows that the ME energies are obtained from the relation
\begin{equation}
\frac{V}{\tau}= \frac{\exp(\lambda)-\exp(-\lambda)}{\exp(M \lambda)-\exp(-M \lambda)}. \label{noo1}
\end{equation}
while for localized eigenstates the Lyapunov exponent reads [Eq.(\ref{uff2})]
\begin{equation}
 \gamma(E)= {\rm log}  \left( \frac{V }{\tau}  \frac{\exp(M\lambda)-\exp(-M\lambda)}{\exp( \lambda)-\exp(-\lambda)} \right). \label{noo2}
 \end{equation}
Taking into account that
\[
\exp (\lambda) \equiv \frac{E}{2 \tau}+ \sqrt{\left( \frac{E}{2 \tau} \right)^2-1} \]
\[ \exp (-\lambda) \equiv \frac{E}{2 \tau}- \sqrt{\left( \frac{E}{2 \tau} \right)^2-1}
\]
from Eqs.(\ref{noo1}) and (\ref{noo2}) we thus retrieve the results of Ref.\cite{M1}.  Note that for this model the $(M-1)$ gaps of the disorder-free superlattice are closed and the $(M-1)$ resonance energies $E=E_{\lambda}$, obtained by letting $\mathcal{K}_{11}(E)=0$, are given by
$E_{\lambda}=2 \tau \cos (\lambda  \pi/ M)$ ($\lambda=1,2,....,M-1$). Since any resonance energy $E_{\lambda}$ belongs to $\sigma(H)$, even in the strong disorder regime, i.e. for large values of potential amplitude $V$, narrow energy regions around $E_{\lambda}$ of extended states arise, i.e. complete Anderson localization is prevented for this model.\\
\\
 \begin{figure*}
    \centering
    \includegraphics[width=0.9 \textwidth]{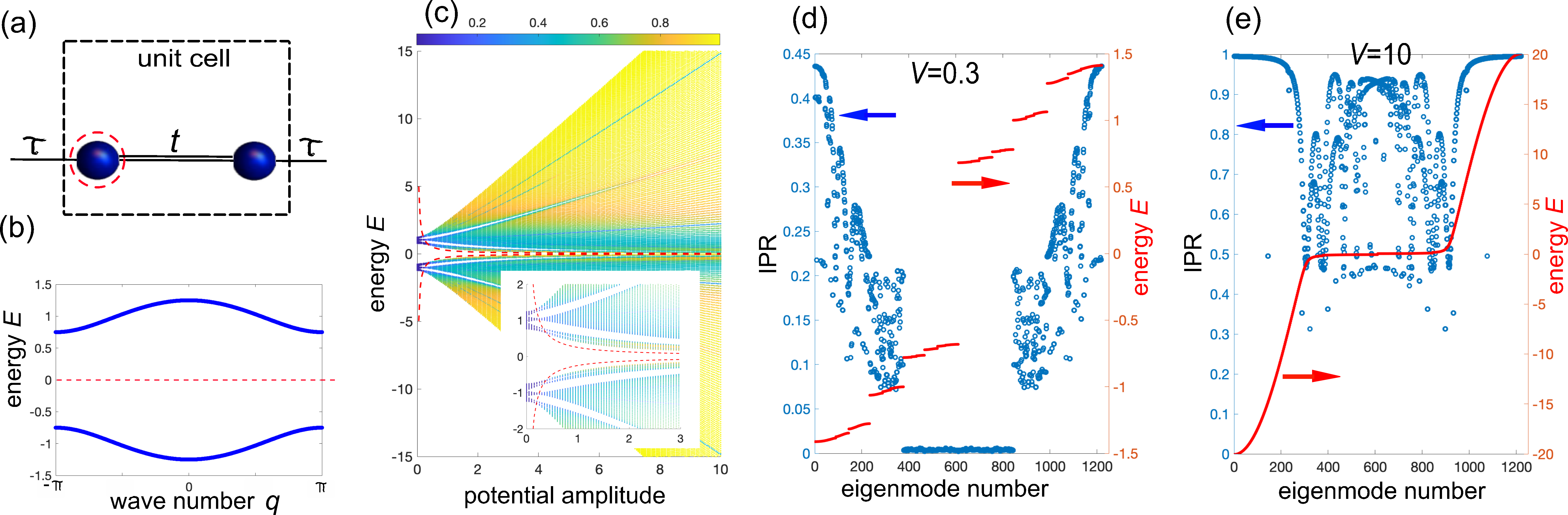}
    \caption{Energy spectra and IPR in a mosaic SSH lattice with incommensurate sinusoidal potential. Parameter values  are $t=1$, $\tau=0.25$, and $\alpha=(\sqrt{5}-1)/2$. In the numerical simulations, the inverse golden ratio $\alpha$ is approximated by the rational number $\alpha \simeq 377/610$, and a finite lattice comprising $L=610$ unit cells with periodic boundary conditions has been assumed. (a) Schematic of the mosaic SSH lattice. (b) Energy spectrum of the disorder-free superlattice. Note that the resonance energy $E_1=0$ falls in the middle of the gap (horizontal dashed line). (c) Numerically-computed energy spectrum and IRP (on a pseudocolor map) versus potential amplitude $V$. The dashed curves are the ME energy curves $E= \pm \tau t/V$ as predicted by Eq.(\ref{mobility}). The inset in the bottom of the figure shows  an enlargement of the plot at low values of the potential amplitude $V$. (d,e) Behavior of the IPR and eigenenergies for all the $2L=1220$ eigenstates of the Hamiltonian corresponding to a disorder strength $V=0.3$ in (d), and $V=10$ in (e). Note that in (e) (strong disorder regime) all eigenstates are exponentially localized, with IPR finite and well above zero for any eigenstates, while in (d) extended (IPR$\sim 0$) and localized states coexist.}
  \end{figure*}
 \begin{figure*}
    \centering
    \includegraphics[width=0.9 \textwidth]{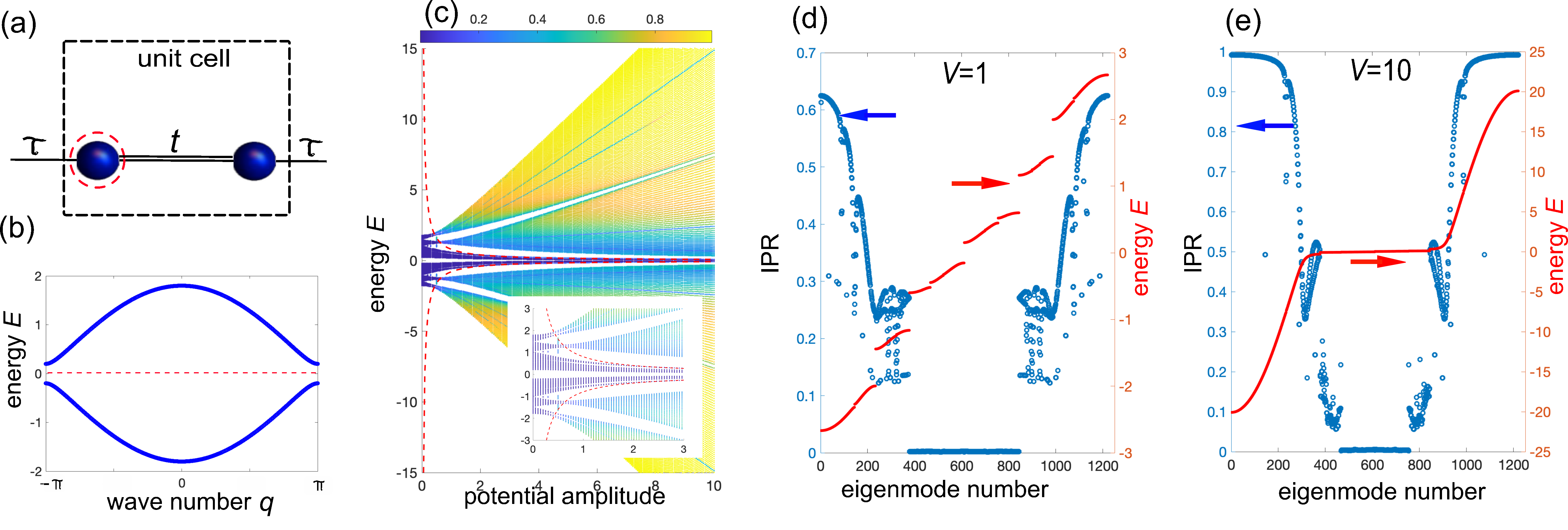}
   \caption{Same as Fig.2, but for $\tau=0.8$, corresponding to a narrow gap. Note that in both  (d) (weak disorder) and (e) (strong disorder) extended and localized states coexist.}
  \end{figure*}
{\it 2) Mosaic Su-Schrieffer-Heeger lattice}. The mosaic SSH model, schematically depicted in Fig.2(a), is a binary lattice ($M=2$) with alternating hopping amplitudes $t$ (intra-cell hopping) and $\tau$ (inter-cell hopping). The incommensurate disorder is applied to one sublattice solely, indicated by the open red circle in Fig.2(a). The energy spectrum of the disorder-free lattice is described y the two minibands
\begin{equation}
\mathcal{E}_{1,2}(q)= \pm \sqrt{\tau^2+t^2+2 \tau t \cos q}
\end{equation}
($-\pi < q \leq \pi$), separated one another by a gap of width $| \tau-t|$ around $E=0$. The gap closes at $\tau=t$, at which the SSH model reduces the the homogeneous lattice model with $M=2$ discussed in the previous example.
The $\mathcal{S}$ matrix for the mosaic SSH model is  given by
\begin{equation}
\mathcal{S}=\left(
\begin{array}{cc}
0 & t \\
t & 0
\end{array}
\right)
\end{equation}
so that the elements $\mathcal{K}_{11}(E)$,  $\mathcal{K}_{22}(E)$  and $\mathcal{K}_{12}(E)$ of the matrix $\mathcal{K}=(E-\mathcal{S})^{-1}$ read
\begin{equation}
\mathcal{K}_{11}(E)=\mathcal{K}_{22}(E)=\frac{E}{E^2-t^2} \; ,\;\; \mathcal{K}_{12}(E)=\frac{t}{E^2-t^2}.
\end{equation}
Note that, unlike the homogeneous lattice limit $t= \tau$ dicsussed in the previous example, the resonance energy for the mosaic SSH model, $E_1=0$, falls in the  gap. According to Eqs.(\ref{uff1}) and (\ref{uff2}), for $|E|> t \tau /V$ the eigenstates are exponentially localized with an inverse localization length $\gamma(E)={\rm ln}  \left\{|E|V /(t \tau) \right\} $ and ME at the energies $E= \pm \tau t/V$ arise.  Note that at strong values of the potential amplitude $V$ the possible extended states have energies in a narrow interval around the resonance energy $E=0$, which shrinks as $V \rightarrow \infty$.  Given that the reonance energy falls in the gap of the disorder-free lattice, an interesting question is whether, in the strong disorder limit $V \gg \tau, t$, all eigenstates in the mosaic SSH are exponentially localized, or extended states can survive for large disorder (as in the limiting case of homogeneous lattice $t=\tau$). To answer this question, let us notice that $E=0$ cannot belong to $\sigma(H)$ because $ |\mathcal{K}_{12}(E=0) \tau| = \tau/t \neq 1$, however small energies close to $E=0$ can. In fact, assuming $E$ small and $V$ large of order $V \sim 1 /E$, the reduced equation (\ref{eq6}) reads
\begin{equation}
\tau t \left( \psi_{n+1}^{(1)}+ \psi_{n-1}^{(1)} \right)+ 2 E V \cos ( 2 \pi \alpha n+\theta) \psi_n^{(1)}= \mathcal{E} \psi_n^{(1)} \label{cazz}
\end{equation}
where
\begin{equation}
\mathcal{E}=-(t^2+ \tau^2).
\end{equation}
An energy $E \in \sigma(H)$ corresponds to an extended state provided that 
\begin{equation}
V|E|< t \tau. \label{cazz2}
\end{equation}
On the other hand, it is clear from Eq.(\ref{cazz}) that $| \mathcal{E}|<2 t \tau+2V|E|$, and thus from Eq.(\ref{cazz2}) the following condition must be necessarily met
\begin{equation}
t^2+ \tau^2 <4 t \tau
\end{equation}
for the eigenenergy $E \in \sigma(H)$ to correspond to an extended state. Therefore, a necessary condition for $H$ to have absolutely continuous spectrum (bands of extended states) in the large potential (disorder) regime is that
\begin{equation}
(2- \sqrt{3}) t< \tau < (2+ \sqrt{3}) t. \label{conditio}
\end{equation} 
When the gap $|t-\tau|$ is sufficiently wide, i.e. the resonance energy $E=0$ is enough far distant from the two energy bands of the disorder-free lattice, the condition (\ref{conditio}) is violated and at strong enough disorder all eigenstates of the mosaic SSH are exponentially localized. In other words, when the resonance energy is deep insight the gap, extended states are prevented at strong enough disorder. This is a rather general result that will be demonstrated for generic mosaic lattice  models in Sec.III.\\ 
Some illustrative examples ef energy spectra and localization properties of the mosaic SSH lattice are shown  in Figs.2 and 3. In the numerical simulations, we assumed the inverse golden mean $\alpha=(\sqrt{5}-1)/2$ as irrational Diophantine. This irrational number is approximated by the sequence of rationals $\alpha= \lim_{n \rightarrow \infty} (F_{n+1}/F_n)$
where $F_n=0,1,1,2,3,5,8,13,21,34,...$ are the Fibonacci
numbers. For finite systems, we assumed a lattice comprising a large number $L=F_n$ unit cells (typically $L=F_{16}=610$) with periodic boundary conditions. Therefore, the total number of lattice sites is $M \times L$. 
The localization properties of the eigenstates $\psi_l$ of $H$ are captured, in the thermodynamic limit $L \rightarrow \infty$,  by  the inverse participation ratio (IPR), which for a normalized wave function is defined as  $\text{IPR} =\sum_{l=1}^{ML} | \psi_l |^{4}$. For localized states,  the IPR takes a finite (and far from zero) value as $L \rightarrow \infty$,  while it assumes a small value, vanishing as $L \rightarrow \infty$, for extended (ergodic or critical) states. Namely, indicating by $\beta= \lim_{L \rightarrow \infty} (  \ln {\rm IPR}) / \ln (1/ML)$ the fractal dimension, for a localized wave function  $\beta=0$, for an extended (ergodic) wave function $\beta=1$, whereas for a critical wave function  $0<\beta<1$. Figures 2(c) and 3(c) show the numerically-computed behavior of the IPR versus potential strength $V$ on a pseudo color map for a SSH lattice with wide gap ($t=1$, $\tau=0.25$, Fig.2) and narrow gap ($t=1$, $\tau=0.8$, Fig.3). As clearly shown in Fig.2(d) and (e), for the wide gap lattice  ME are observed at low disorder strength $V$, while all eigenstates become localized for large disorder. Conversely, in the narrow gap lattice ME persist also for strong disorder, with extended states at energies close to the resonance energy $E=0$, as shown in Figs.3(d) and (e).\\
\\ 
{\it 2. Mosaic trimer lattice}. The mosaic trimer lattice is schematically shown in Fig.4(a). Each unit cell comprises three sites ($M=3$) with intra-cell hopping amplitudes $t_1$, $t_2$ and $t_3$, which are assumed to be real positive numbers. For the disorder-free lattice, the dispersion curves $\mathcal{E}_{\lambda}(q)$ of the three superlattice minibands are obtained as the eigenvalues of the matrix $\mathcal{A}$ [Eq.(\ref{matrixA})] and satisfy the cubic equation
\begin{equation}
\mathcal{E}^3+u(q)\mathcal{E}+v(q)=0
\end{equation}
which can be solved using Cardano's formula
\begin{equation}
\mathcal{E}_{\lambda}(q)=2 \sqrt{-\frac{u}{3}} \cos \left[ \frac{1}{3} {{\rm acos} \left( \frac{3v}{2u} \sqrt{-\frac{3}{u}} \right)} + \frac{2 \pi}{3} \lambda \right]
\end{equation}
with $\lambda=1,2,3$ (band index) and $\mathcal{E}_1(q) \leq \mathcal{E}_2(q) \leq \mathcal{E}_3(k)$). In the above equations we have set
\begin{eqnarray}
u(q) & = & -(t_1^2+t_2^2+t_3^2+\tau^2+2 \tau t_3 \cos q) \\
v(q) & = & -2(t_1t_2t_3+\tau t_1 t_2 \cos q).
\end{eqnarray}
Typical energy band diagrams of the disorder-free lattice are shown in Figs.4(b) and 5(b). The intra-cell matrix Hamiltonian $\mathcal{S}$ for the trimer superlattice reads
   \begin{figure*}
    \centering
    \includegraphics[width=0.9 \textwidth]{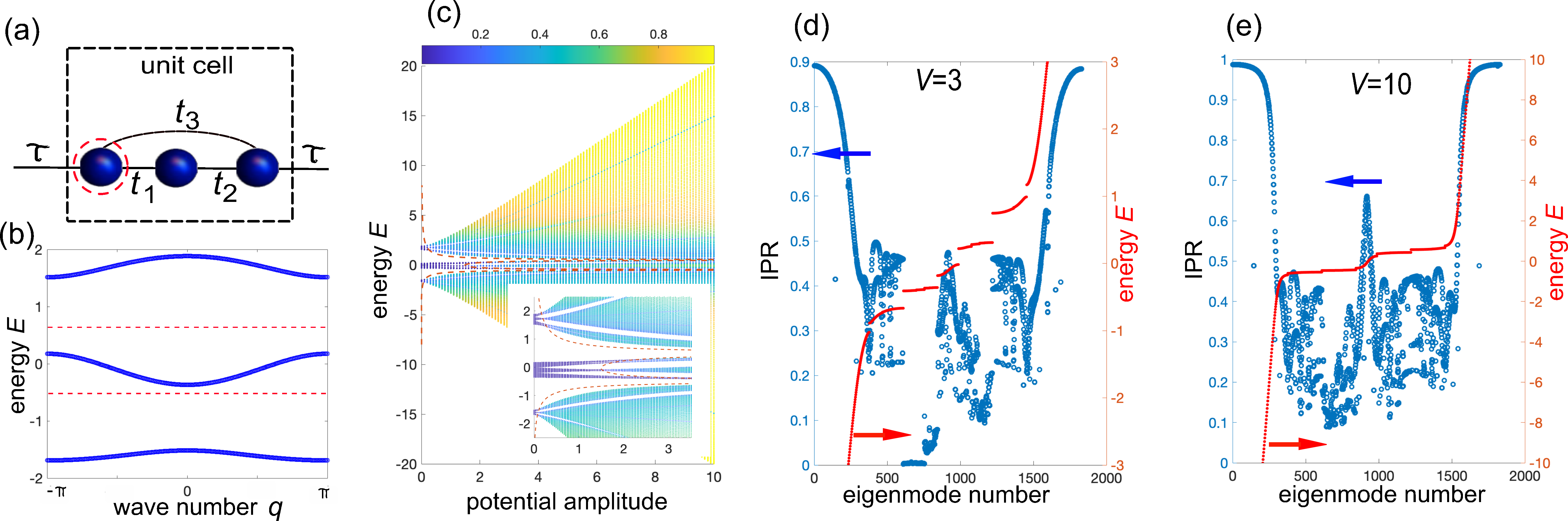}
    \caption{Energy spectra and IPR in a mosaic trimer lattice with incommensurate sinusoidal potential. Parameter values  are $t_1=1.5$, $t_2=0.5$, $t_3=0.2$, $\tau=0.5$ and $\alpha=(\sqrt{5}-1)/2$. In the numerical simulations, the inverse golden ratio $\alpha$ is approximated by the rational number $\alpha \simeq 377/610$, and a finite lattice comprising $L=610$ unit cells with periodic boundary conditions has been assumed. (a) Schematic of the mosaic lattice. (b) Energy spectrum of the disorder-free superlattice. Note that the two resonance energies $E_1=-t_2$ and $E_2=t_2$, depicted by the two horizontal dashed lines, fall in the gaps. (c) Numerically-computed energy spectrum and IRP (on a pseudocolor map) versus potential amplitude $V$. The dashed curves are the ME energy curves as predicted by Eq.(\ref{mobility}). The inset in the bottom of the figure shows  an enlargement of the plot at low values of the potential amplitude $V$. (d,e) Behavior of the IPR and eigenenergies for all the $3L=1830$ eigenstates of the Hamiltonian corresponding to a disorder strength $V=3$ in (d), and $V=10$ in (e). Note that in (e) (strong disorder regime) all eigenstates are exponentially localized, with IPR finite and well above zero for any eigenstates, while in (d) extended and localized states coexist.}
  \end{figure*}
    \begin{figure*}
   \centering
    \includegraphics[width=0.9 \textwidth]{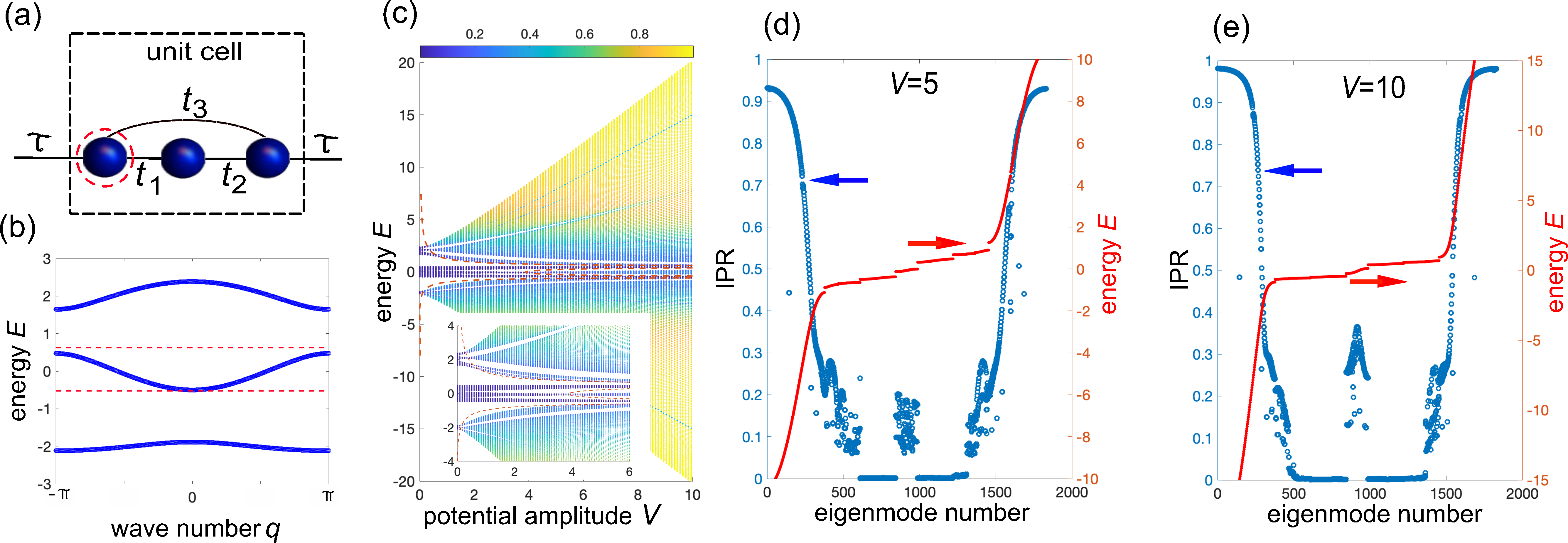}
   \caption{Same as Fig.4, but for $\tau=|t_1-t_3|=1.3$. In this case one of the two resonance energies, namely $E_1=-t_2$, is at the bottom band edge of the middle band, as shown in panel (b). As compared to Fig.4, even in the strong disorder regime $V=10$ [panel(e)] a large fraction of extended states coexist with localized states.}
  \end{figure*}
\begin{equation}
\mathcal{S}=
\left( 
\begin{array}{ccc}
0 & t_1 & t_3 \\
t_1 & 0 & t_2 \\
t_3 & t_2 & 0
\end{array}
\right)
\end{equation}
from which the elements $\mathcal{K}_{11}$, $\mathcal{K}_{13}$ and $\mathcal{K}_{33}$ of the matrix $\mathcal{K}=(E- \mathcal{S})^{-1}$ can be readily obtained
\begin{eqnarray}
\mathcal{K}_{11} & = & \frac{E^2-t_2^2}{E^3-(t_1^2+t_2^2+t_3^2)E-2t_1t_2t_3} \\
\mathcal{K}_{13} & = & \frac{t_1t_2+Et_3}{E^3-(t_1^2+t_2^2+t_3^2)E-2t_1t_2t_3}  \\
\mathcal{K}_{33} & = & \frac{E^2-t_1^2}{E^3-(t_1^2+t_2^2+t_3^2)E-2t_1t_2t_3} 
\end{eqnarray}
The two resonance energies $E_{\lambda}$, i.e. the roots of $\mathcal{K}_{11}(E)=0$, are thus given by $E_{1}=-t_2$ and $E_2=t_2$. The energy $E_{\lambda}$ belongs to the spectrum $\sigma(H)$ whenever $|\mathcal{K}_{13}(E_{\lambda}) \tau|=1$. Specifically, the resonance energy $E_1=-t_2 \in \sigma (H)$ when $\tau=|t_1-t_3|$, whereas  the resonance energy $E_2=t_2 \in \sigma (H)$ when $\tau=t_1+t_3$. According to Eq.(\ref{mobility}), for this model the ME are found from the relation
\begin{equation}
V= \tau \left|  \frac{t_1t_2+Et_3}{E^2-t_2^2} \right|.
\end{equation}
Two examples of ME and localization properties of the mosaic trimer lattice are shown in Figs.4 and 5. In Fig.4 the two resonance energies fall deep inside the two gaps of the superlattice, such that at strong enough values of $V$ all eigenstates are exponentially localized, as shown in Fig.4(e), Conversely, in Fig.5 the resonance energy $E_1=-t_2$ is at the bottom edge of the middle band, so that a non-negligible fraction of extended states, with energies close to $E_1$, are found also in the strong disorder regime, as clearly shown in Fig.5(e).  

\subsubsection{Random disorder}
In this case the mosaic on-site potentials $V_n$ are assumed to be independent random variables with the same probability density distribution $f(V)$ of zero mean. For $M=1$, it is well known that for any arbitrarily small (weak) disorder all eigenstates of $H$ are exponentially localized with Lyapunov exponent $\gamma(E)$ strictly positive at any energy
\cite{R2bis,R2b,R2c,S3c}, with ${\rm Inf}_{E \in \sigma(H)} \gamma(E) >0$ \cite{S3c}.\\
 For $M \geq 2$, we should distinguish two cases.\\ 
If one (or more) resonance energies $E_{\lambda}$ falls in an energy band of the disorder-free lattice, then $E_{\lambda} \in \sigma(H)$ and the corresponding eigenstate is an extended state, so that complete Anderson localization is prevented even in the strong disorder regime. As already discussed in previous works \cite{M2,M3,D1,D2,D3,D4,D5}, the density of states $\rho(E)$ and localization length $ 1 / \gamma(E)$ show characteristic resonance peaks (singularities) at $E=E_{\lambda}$. In fact, for an energy $E \sim E_{\lambda}$ close to a resonance, taking into account that $\tau \mathcal{K}_{1M}(E)= \pm 1+O(\epsilon) $ and $\mathcal{K}_{11}(E) \sim \epsilon$ with $\epsilon \equiv (E-E_{\lambda})$, at leading order in $\epsilon$ Eq.(\ref{eq6}) yields
\begin{equation}
\pm ( \psi_{n+1}^{(1)}+\psi_{n-1}^{(1)})+W_n \psi_n{(1)}= 2 \psi_n^{(1)}
\end{equation}
 where $W_n=\mathcal{K}_{11}(E) V_n \sim \epsilon V_n$ is the effective random disorder.  Since such an effective disorder is weak (of order $\sim \epsilon$), for any probability distribution $f(V)$ with finite variance $\langle V^2 \rangle$ the following asymptotic scaling holds for the inverse localization length $\gamma(E)$ and density of states $\rho(E)$ \cite{Derrida}
 \begin{equation}
 \gamma(E) \sim \epsilon^{2/3 } \langle V^2 \rangle^{1/3}  \; ,\; \; \rho(E) \sim  \epsilon^{-2/3 } \langle V^2 \rangle^{-1/3} \label{density}
 \end{equation}
 with $\epsilon=E-E_{\lambda}$. Hence a set of exponentially-localized eigenstates with diverging localization length accumulate as the energy $E$ approaches the resonance energy $E_{\lambda}$.\\
 Conversely, if all resonance energies $E_{\lambda}$ fall in the gaps of the disorder-free lattice, then $E_{\lambda} \notin \sigma(H)$ and all eigenstates of $H$ are exponentially localized even in the weak disorder regime. In fact, according to Eq.(\ref{eq6}) for any energy $E$ different than $E_{\lambda}$  the effective random disorder   $W_n=\mathcal{K}_{11}(E) V_n$ is non-vanishing, implying Anderson localization. Also, as shown in the next section (Theorem 2), ${\rm inf}_{E \in \sigma(H)} \{ \gamma(E) \}>0$, i.e. there is a finite  (non-diverging) upper bound for the localization length $ 1 / \gamma(E)$ of all exponentially-localized eigenstates.\\ 
\\
The above results can be illustrated by considering an exactly-solvable model of Anderson localization in one-dimension, the Lloyd model \cite{Lloyd} (see also \cite{R2bis,S3b,BS}), where the probability density distribution $f(V)$ is given by the Cauchy distribution
\begin{equation}
f(V)=\frac{\delta}{\pi} \frac{1}{V^2+\delta^2}
\end{equation}
where $\delta$ is the width of the distribution. Note that such a distribution does not have any finite moment, so the scaling Eq.(\ref{density}) cannot be used here. Using the effective equation (\ref{eq6}) and the analytical form of the Lyapunov exponent for the Lloyd model \cite{R2bis,S3b}, for a generic mosaic random lattice one obtains the following exact result
\begin{widetext}
\begin{equation}
\gamma(E)= {\rm acosh} \left(  \frac{\sqrt{[2+\mathcal{L}(E)]^2+R^2(E)} + \sqrt{[2-\mathcal{L}(E)]^2+R^2(E)}}{4}  \right) \label{Lia}
\end{equation}
\end{widetext}
for the Lyapunov exponent, where we have set
\begin{eqnarray}
\mathcal{L}(E) & = & \frac {  1+ \tau^2 |\mathcal{K}_{1M}(E) |^2 -\tau^2 \mathcal{K}_{11} (E) \mathcal{K}_{MM}(E)} { \tau | {\mathcal{K}_{1M}(E)|}} \label{palle1} \\
R(E) & = & \delta \left| \frac{\mathcal{K}_{11}(E)}{\mathcal{K}_{1M}(E) } \right| \label{palle2}.
\end{eqnarray}
   \begin{figure}
    \centering
   \includegraphics[width=0.5 \textwidth]{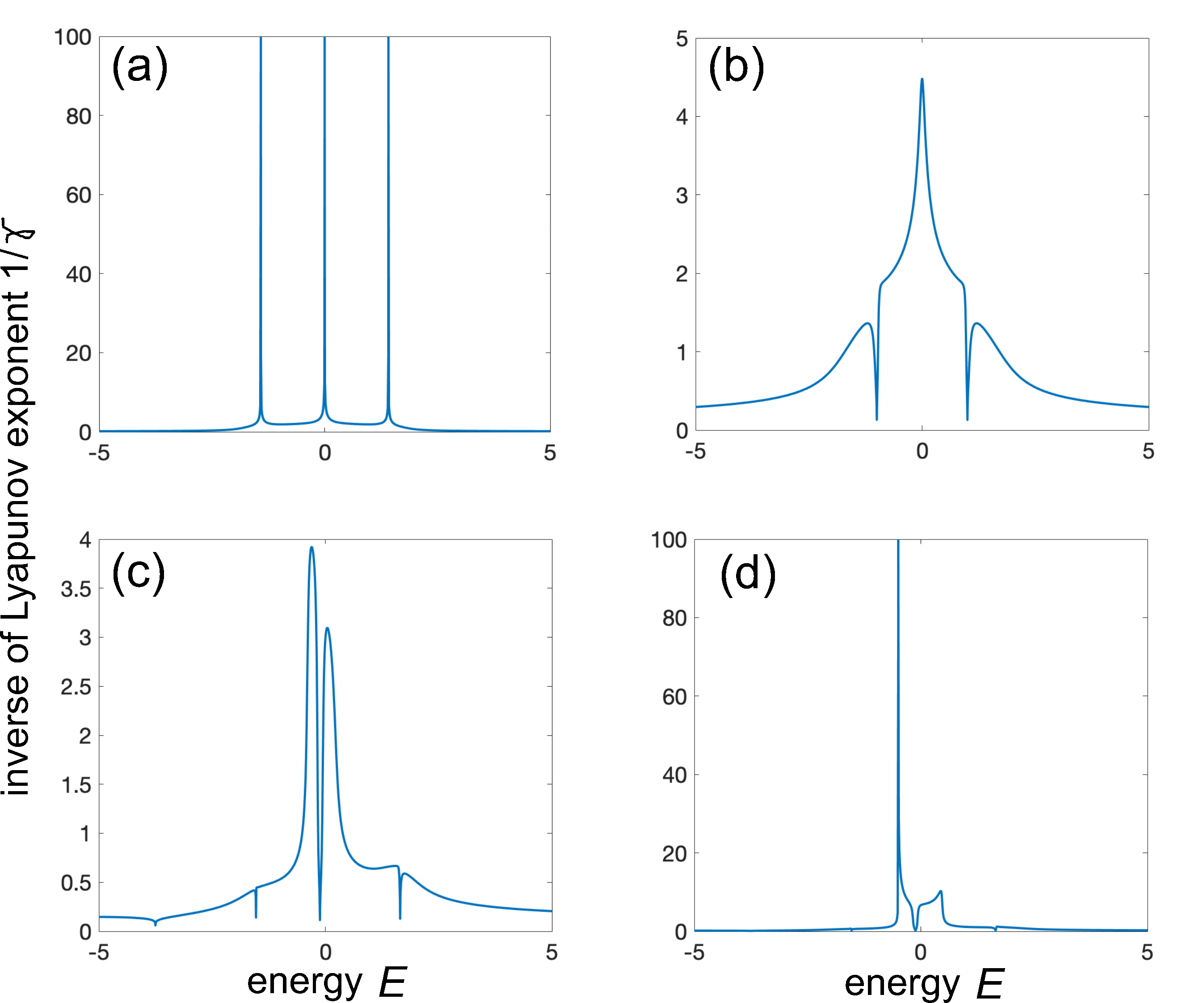}
   \caption{Behavior of the inverse of Lyapunov exponent $1 / \gamma(E)$ (localization length) versus energy $E$ for random uncorrelated disorder described by a Cauchy distribution of width $\delta=1$ in four different mosaic lattice models. (a) Homogeneous lattice with nearest-neighbor hopping amplitude $\tau=1$ and $M=4$. The disorder-free model is gapless and at the three resonance energies $E_{\lambda}=2 \tau \cos (\lambda  \pi/ 4)$ ($\lambda=1,2,3$) the localization length diverges (extended states belonging to $\sigma(H)$). (b) SSH mosaic lattice with $t=1$ and $\tau=0.8$. The model displays a resonance energy $E_1=0$ in the gap, and all eigenstates are exponentially localized. The energy spectrum of $H$ is the entire energy axis, with the exception of $E=E_1=0$. (c) Mosaic trimer model with $t_1=1.5$, $t_2=0.5$, $t_3=0.2$ and $\tau=0.5$. The two resonance energies $E_{1,2}= \pm t_2$ are in the gaps of the disorder-free lattice [see Fig.4(b)] and all eigenstates are exponentially localized. (d) Same as (c), but for $\tau=|t_1-t_3|=1.3$. In this case the resonance energy $E_1=-t_2$ is at the bottom edge of the middle band of the disorder-free lattice [see Fig.5(b)], and the localization length diverges at such an energy.}
  \end{figure}
 Extended states, if any, correspond to $\gamma(E)=0$, whereas $\gamma(E)>0$ corresponds to either $E \in \sigma(H)$ and exponentially-localized eigenstate, with localization length $1 / \gamma(E)$, or to an energy $E$ not in the spectrum of $H$. It can be readily shown that, as it should be, for any $\delta>0$ one has $\gamma(E)=0$ if and only  if $E$ is one of the resonance energies $E_{\lambda}$: in this case they belong to the spectrum of $H$. In fact, from Eq.(\ref{Lia}) it follows that $\lambda(E)$ vanishes if and only if $R(E)=0$ {\em} and $|\mathcal{L}(E)| \leq 2$, which using Eqs.(\ref{palle1}) and (\ref{palle2}) necessarily implies $\mathcal{K}_{11}(E)=0$  and $\tau | \mathcal{K}_{1M}(E)|=1$, i.e. $E=E_{\lambda}$ and $E_{\lambda} \in \sigma(H)$. Moreover, since the density of states of the Lloyd model \cite{Lloyd,BS} is nonvanishing and regular at any energy,  it follows that for any $\delta>0$ one has $\sigma(H)=\{-\infty < E < \infty\}$, with the exclusion of some isolated points, such as the resonances $E=E_{\lambda}$ if $ \gamma (E_{\lambda}) >0$. Figure 6 shows typical behavior of $\gamma(E)$ for the same three illustrative models previously introduced for the incommensurate disorder case: the nearest-neighbor lattice with uniform hopping amplitude $\tau$ [Fig.6(a)], the SSH mosaic lattice [Fig.6(b)], and the trimer lattice [Figs.6(c) and (d)]. Note that in Fig.6(a) and (d) at least one resonance frequency falls in a band of the disorder-free lattice, the eigenstates are extended and correspondingly the localization length diverges at such energies. Conversely, in Figs.6(b) and (c) the resonance energies $E_{\lambda}$ fall in the gaps of the disorder-free lattice, $\gamma(E_{\lambda})>0$ and such energies do not belong to the spectrum of $H$.

 \section{Gap-protected Anderson localization}
 The illustrative models considered in previous section indicate that, when all resonance frequencies fall deeply in the gaps of the disorder-free superlattice, all eigenstates are exponentially localized in the strong disorder regime (for incommensurate disorder) or even for any small disorder strength (for uncorrelated disorder).  Here we derive some general criteria under which in a generic mosaic lattice, with either incommensurate (sinusoidal) or uncorrelated disorder, all eigenstates are exponentially (Anderson) localized. To this aim, we first introduce a useful Lemma, which provides a necessary and sufficient condition for the resonance energies to fall in the gaps of the disorder-free superlattice:\\
 {\em Lemma 1.} Let us consider the mosaic lattice described by the Hamiltonian $H$ as in Eq.(\ref{eq2}), with inter-cell hopping amplitude $\tau$ and intra-cell Hamiltonian $\mathcal{S}$ [Fig.1(a)]. Let $\mathcal{K}(E)=(E-\mathcal{S})^{-1}$ and $E_{\lambda}$ ($ \lambda=1,2,...,M-1$) the resonance energies, corresponding to the $(M-1)$ roots of $\mathcal{K}_{11}(E)=0$. 
  Then the resonance energies are strictly inside the gaps of the disorder-free lattice, i.e.
\begin{equation}
\mathcal{E}_1(q) < E_1 < \mathcal{E}_2(q) < E_2 <... <E_{M-1} < \mathcal{E}_{M}(q)
\end{equation} 
if and only if all the $(M-1)$ gaps of the disorder-free lattice are open and if $| \tau \mathcal{K}_{1M}(E_{\lambda})| \neq 1$ for any $\lambda$. \\
The Lemma can be proven using the Cauchy's interlace theorem for the eigenvalues of Hermitian matrices \cite{inter}, which yields the inequalities Eq.(\ref{ineq}). Further, since $| \tau \mathcal{K}_{1M}(E_{\lambda})| \neq 1$, $E_{\lambda}$ cannot belong to the spectrum $\sigma(H_0)$, and thus the inequalities in Eq.(\ref{ineq}) are strict (i.e. $<$ rather than $\leq$).

 \subsection{ Incommensurate (sinusoidal) disorder}
 The following theorem provides a sufficient condition for gap-protection of Anderson localization in the case of a mosaic lattice with sinusoidal incommensurate potential, i.e. of the Aubry-Andr\'e (or almost Mathieu) type [Eq.(\ref{poten})].\\
 {\em Theorem 1.} Let us consider a generic mosaic lattice, as described by the Hamiltonian $H$ in Eq.(\ref{eq2}), with an incommensurate sinusoidal potential $V_n= 2 V \cos ( 2 \pi \alpha n+ \theta)$ ($ \alpha$ irrational Diophantine). Let us assume that the resonance energies $E_{\lambda}$ fall in the gaps of the disorder-free superlattice, i.e. that the conditions given in Lemma 1 are satisfied. If the   resonance energies are deep enough in the gaps, namely if
 \begin{equation}
 \tau |\mathcal{K}_{1M}(E_{\lambda})| > 2+ \sqrt{3} \; \; {\rm or} \;\;\ \tau |\mathcal{K}_{1M}(E_{\lambda})| < 2- \sqrt{3}  \label{condAnder}
 \end{equation}
 for any $\lambda=1,2,...,M-1$, then there exists a value $V=V^{(m)}$ of the potential amplitude such that for $V>V^{(m)}$ all eigenstates of $H$ are exponentially localized.\\
  {\em Proof.}  Since $\mathcal{K}(E)$ is a continuous (non-singular) function of energy $E$ near the resonance energies, we can consider a small (yet finite) energy interval $\Delta E$, such that for any energy $E=E_{\lambda}+ \epsilon$ ($| \epsilon | < \Delta E$)  the elements of $\mathcal{K}(E)$ can be expanded in power series in $\epsilon$ up to $\sim \epsilon$. Indicating by $\mathcal{P}$ the set of energy intervals defined by $\mathcal{P}=\bigcup_{\lambda=1}^{(M-1)} \{ E : \; |E-E_{\lambda}|< \Delta E \}$, for any energy $E$ in the spectrum of $H$ but not belonging to the set $\mathcal{P}$, i.e. for any $E \in \sigma(H) { \setminus }  \mathcal{P}$, the quantity $| \mathcal{K}_{1M}(E)  /  \mathcal{K}_{11}(E)|$ is bounded from above, i.e. ${\rm Sup}_{E \in \sigma(H) { \setminus}  \mathcal{P}}| \mathcal{K}_{1M}(E) / \mathcal{K}_{11}(E) |< \infty$, because the poles of $\mathcal{K}_{1M}(E) / \mathcal{K}_{11}(E)$ exactly arise at $E=E_{\lambda}$, which are excluded from the set $\sigma(H) {\setminus} \mathcal {P}$. If we assume a potential amplitude $V=V^{(m)}$ given by
   \begin{equation}
 V^{(m)} =  {\rm Sup}_{E \in \sigma(H) {\setminus} \mathcal{P}}\left|  \tau \frac{\mathcal{K}_{1M}(E)}{\mathcal{K}_{11}(E)} \right|
 \end{equation}
 from Eq.(\ref{uff1}) it follows that for any energy $E \in \sigma(H) \setminus \mathcal{P}$ the corresponding eigenfunction is exponentially localized. To prove the theorem, it is enough to show that for any energy $E \in \sigma(H) \bigcap \mathcal{P}$ the eigenstates of $H$ cannot be extended (ergodic or critical) states. In fact, for $E \in \sigma(H) \bigcap \mathcal{P}$ at leading order in $\epsilon=E-E_{\lambda}$ the reduced equation  (\ref{eq6}) reads
 
   \begin{eqnarray}
 \tau \mathcal{K}_{1M}(E_{\lambda}) \psi_{n+1}^{(1)}+\tau \mathcal{K}_{M1}(E_{\lambda}) \psi_{n-1}^{(1)}+ \nonumber \\ 
 \mathcal{K}_{11}(E_{\lambda}+ \epsilon)  V_n  \psi_{n}^{(1)}
 = \mathcal{E}  \psi_{n}^{(1)} \label{figa1}
 \end{eqnarray}
 with  $ \mathcal{E}=1+\tau^2 |\mathcal{K}_{1M}(E_{\lambda})|^2$. Clearly, the following  bound for the eigenvalue $\mathcal{E}$ of Eq.(\ref{figa1})
 \begin{equation}
 | \mathcal{E}|< 2 \tau |\mathcal{K}_{1M}(E_{\lambda})|+2V^{(m)} |\mathcal{K}_{11}(E_{\lambda}+ \epsilon) |,
 \end{equation}
 holds \cite{nota}, and thus one has
 \begin{equation}
1+\tau^2 |\mathcal{K}_{1M}(E_{\lambda)}|^2< 2 \tau |\mathcal{K}_{1M}(E_{\lambda})|+2V^{(m)} |\mathcal{K}_{11}(E_{\lambda}+ \epsilon) |. \label{figa2}
 \end{equation}
 On the other hand, if the eigenstate with energy $E=E_{\lambda}+ \epsilon$ were extended (ergodic or critical), 
 according to Eq.(\ref{extended}) one should have $V^{(m)} \leq \tau | \mathcal{K}_{1M}(E_{\lambda}) / \mathcal{K}_{11}(E_{\lambda}+ \epsilon)|$, so that from Eq.(\ref{figa2}) one obtains
 \begin{eqnarray}
 1+\tau^2 |\mathcal{K}_{1M}(E_{\lambda)}|^2  & < &  2 \tau |\mathcal{K}_{1M}(E_{\lambda})|+2V^{(m)} |\mathcal{K}_{11}(E_{\lambda}+ \epsilon) |  \nonumber \\
 & < &  4 \tau |\mathcal{K}_{1M}(E_{\lambda})|, \nonumber
 \end{eqnarray}
 i.e.
  \begin{equation}
 \tau^2 |\mathcal{K}_{1M}(E_{\lambda)}|^2  - 4 \tau |\mathcal{K}_{1M}(E_{\lambda})|+1<0 \label{figa3}
 \end{equation}
 for the eigenstate to be extended. However, when the inter-hopping amplitude $\tau$ is in the range defined by Eq.(\ref{condAnder}), it can be readily shown that Eq.(\ref{figa3}) is never satisfied, i.e. there are not extended states in the spectral range $E \in \sigma(H) \bigcap \mathcal {P}$. This concludes the proof of the theorem.\\
 {\em Remark.} Physically, the condition (\ref{condAnder}) means that the resonance energies $E_{\lambda}$ should be  deeply insight the gaps of the disorder-free lattice, i.e. far enough from the band edges. In fact, in the disorder-free lattice the Lyapunov exponent $\gamma(E)$ vanishes if and only if $E \in \sigma(H_0)$, i.e. if the energy $E$ is inside one of the allowed minibands of the superlattice, whereas $\gamma(E)>0$ if the energy $E$ is not in the spectrum, i.e. it is in a gap. Roughly speaking, $\gamma(E)$ measures the 'distance' of the energy $E$ from band edges: the further the Lyapunov exponent $\gamma(E)$ is far apart from zero, the further the energy $E$ is from the edges of the allowed bands. For $E= E_{\lambda}$ the Lyapunov exponent is given by 
 \begin{equation}
 \gamma(E_{\lambda})=\left|  {\rm ln}  \left|  \tau \mathcal{K}_{1M} (E_{\lambda}) \right| \right|
 \end{equation}
  which is readily obtained from Eq.(\ref{figa1}) by letting $\epsilon \rightarrow 0$.   
  The condition (\ref{condAnder}) is equivalent to state $\gamma(E_{\lambda})> {\rm ln}(2+ \sqrt{3}) \simeq 1.317$, indicating a minimum 'distance' of the resonance energies $E_{\lambda}$ from the band edges. 
 
 \subsection{ Uncorrelated random disorder}
 The following theorem provides a necessary and sufficient condition for gap-protection of Anderson localization in the case of a mosaic lattice with uncorrelated random disorder.\\
 {\em Theorem 2.} Let us consider a generic mosaic lattice, as described by the Hamiltonian $H$ in Eq.(\ref{eq2}), where $V_n$ are uncorrelated random variables with the same probability density distribution $f(V)$. For any arbitrary disorder strength all eigenstates of $H$ are (almost surely) exponentially localized if and only if  the resonance energies $E_{\lambda}$ fall in the gaps of the disorder-free superlattice, i.e. if and only if the conditions given in Lemma 1 are satisfied. In this case there is an upper bound for the localization length of the eigenstates, i.e. 
 \begin{equation}
 {\rm Sup}_{E \in \sigma(H)} \frac{1}{\gamma(E)} < \infty.
 \end{equation}
 {\em Proof.} The condition is obviously necessary: if a resonance energy $E_{\lambda}$ is inside a band (or at the band edge) of the disorder-free superlattice, then $E_{\lambda} \in \sigma(H)$ as well, the eigenstate at $E=E_{\lambda}$ is extended and $\gamma(E_{\lambda})=0$. To prove that the condition is also sufficient, let us remind a basic result of  Anderson localization in one-dimensional lattices with uniform nearest-neighbor hopping and uncorrelated on-site potential disorder: for any arbitrarily weak disorder strength, $\gamma(E)>0$ at any energy and (almost surely) all eigenstates are exponentially localized with ${\rm Inf}_E \gamma(E)>0$  (see e.g. \cite{R2bis,S3c}). Since $E=E_{\lambda}$ does not belong to the spectrum of $H$, for any $E \in \sigma(H)$ one has $\mathcal{K}_{11}(E) \neq 0$, and thus the strength of the random potential $W_n=\mathcal{K}_{11}(E) V_n$ entering in the effective Anderson model (\ref{eq6}) is non-vanishing, implying $\gamma(E)>0$. To prove that there is an upper bound to the localization length $1 / \gamma(E)$ of eigenstates as $E$ varies in $\sigma(H)$, let us notice that since $\gamma(E)$ is a continuous function of $E$ \cite{S3c} and $\gamma(E_{\lambda})=| {\rm ln}  |\tau \mathcal{K}_{1M}(E_{\lambda})| |>0$, there exists a small range $\Delta E$ of energies such as, for any $E$ in the set $\mathcal{P}=\{ E: \; |E-E_{\lambda}|< \Delta   E \} $ one has $\gamma(E)> 1/ \Theta_1$, where $ 1/ \Theta_1 \equiv 
(1/2)  {\rm min}_{\lambda} \gamma(E_{\lambda})$. This means that
\begin{equation}
{\rm Sup}_{E \in \sigma (H) \bigcap \mathcal{P}} \left( \frac{1}{\gamma(E)}  \right)< \Theta_1.
\end{equation} 
On the other hand, for any energy $E \in \sigma(H) \setminus \mathcal{P}$ one has ${\rm Inf}_{E} | \mathcal{K}_{11}(E)|>0$ and thus ${\rm Inf}_E \gamma(E) >0$ (according to theorem 2.6 in \cite{S3c}). After letting ${\rm Inf}_{E \in \sigma(H) \setminus \mathcal{P}} ( \gamma(E)) = 1 / \Theta_2$ and indicating by $\Theta$ the largest value between $\Theta_1$ and $\Theta_2$, one finally obtains
\begin{equation} 
{\rm Sup}_{E \in \sigma(H)} \left(  \frac{1}{\gamma(E)} \right)< \Theta
\end{equation}
 which completes the proof of the theorem. 
  
  \section{Conclusion and discussion}
In summary, we unveiled the general localization properties of a broad class of one-dimensional disordered mosaic lattices, with either incommensurate or uncorrelated disorder, beyond models so far investigated in the open literature \cite{M1,M2,M3,M4,M5,
M7,M8,M9,M10,M11,D1,D2,D3,D4}, extending the block decimation scheme earlier introduced in \cite{D1,D2}.  
 For incommensurate (sinusoidal) disorder, we provided general analytical form of mobility edges and inverse localization length, without resorting to Avila's global theory which can be applied to simple mosaic models solely \cite{M1}. For uncorrelated  disorder, general asymptotic forms of Lyapunov exponent and density of states near resonance energies have been derived, and exact results have been presented  for the class of mosaic Lloyd models. Finally, two general criteria for complete Anderson localization in mosaic lattices have been demonstrated. Specifically, for incommensurate disorder we showed that a sufficient condition for complete Anderson localization at strong enough disorder is that all the gaps of the disorder-free superlattice are open and  the resonance energies fall deeply insight into the gaps. For random disorder, complete Anderson localization is obtained at any disorder strength if and only if all the gaps are open and the resonance energies are not at the band edges.\\
  Our study extends to a broad class of disordered mosaic lattices previous results on mobility edges and emergent resonance energies, unraveling the key role of open gaps of the disorder-free lattice to protect Anderson localization. 
It should be mentioned that the main results obtained in the present work, such as the gap-protection criteria demonstrated in Sec.III, assume nearest-neighbor hopping between adjacent cells of the mosaic superlattice, i.e. models with inter-cell long-range hopping are excluded from the analysis. For such models, unfortunately the block decimation scheme cannot be applied in a simple and general manner, motivating the search for new ways to approach the localization properties of mosaic lattices accounting for inter-cell long-range hopping.

  \section{Acknowledgments}
The author acknowledges the Spanish State Research Agency, through the Severo Ochoa
and Maria de Maeztu Program for Centers and Units of Excellence in R\&D (Grant No. MDM-2017-0711).


\begin{thebibliography}{99}
\bibitem{footzero}
When $E$ is one of the $M$ eigenvalues $\overline{E}_{\lambda}$ of the intra-cell matrix Hamiltonian $\mathcal{S}$, the analysis is still valid provided that the limit $E \rightarrow \overline{E}_{\lambda}$ is made in Eqs.(\ref{eq6}) and (\ref{eq7}). Basically, in such equations the dominant terms arising from the poles of the elements $\mathcal{K}_{1M}(E)$, $\mathcal{K}_{MM}(E)$ and $\mathcal{K}_{11}(E)$ 
should be kept. Rather generally, whenever the rational functions $\mathcal{K}_{1M}(E)$, $\mathcal{K}_{MM}(E)$ and $\mathcal{K}_{11}(E)$ have a pole of the same order at $E= \overline{E}_{\lambda}$, i.e. there is not any accidental cancellation of the pole with a zero in the numerator of the functions, from Eqs.(\ref{eq6}) and (\ref{eq7}) is readily follows that $\mathcal{E}$ diverges faster than  the hopping amplitude $\mathcal{K}_{1M}(E) \tau$ and $\mathcal{K}_{11}(E)$ near the pole, so that for any bounded potential $V_n$ it follows that $E=\overline{E}_{\lambda}$ does not belong  to $\sigma(H)$.
\bibitem{nota}
This bound follows from the very general property that the eigenvalues $\mathcal{E}$ of the lattice equation $\psi_{n+1}+\psi_{n-1}+W_n \psi_n=\mathcal{E} \psi_n$ with a bounded potential $|W_n| < W$ satisfy the condition $| \mathcal{E} |<2+W$.
\end{thebibliography}
\end{document}